\documentclass{WileyMSP-template}
\usepackage{braket} 
\usepackage{verbatim}
\usepackage{xcolor}
\usepackage{float}
\usepackage{soul}
\usepackage{physics}
\usepackage{mathtools}
\usepackage{amsfonts}
\usepackage{amsmath}
\usepackage{dsfont}
\usepackage{cite}
\begin{document}


\title{Benefits of Open Quantum Systems for Quantum Machine Learning}

\maketitle



\author{Mar\'{i}a Laura Olivera-Atencio}
\author{Lucas Lamata}
\author{Jes\'us Casado-Pascual*}


\begin{affiliations}
Mar\'{i}a Laura Olivera-Atencio\\
F\'{\i}sica Te\'orica, Universidad de Sevilla, Apartado de Correos 1065, Sevilla 41080, Spain\\
Lucas Lamata\\
Departamento de F\'isica At\'omica, Molecular y Nuclear, Universidad de Sevilla, 41080 Sevilla, Spain\\
Instituto Carlos I de F\'isica Te\'orica y Computacional, 18071 Granada, Spain\\
Jes\'us Casado-Pascual\\
F\'{\i}sica Te\'orica, Universidad de Sevilla, Apartado de Correos 1065, Sevilla 41080, Spain
\end{affiliations}



\begin{abstract}
Quantum machine learning is a discipline that holds the promise of revolutionizing data processing and problem-solving. However, dissipation and noise arising from the coupling with the environment are commonly perceived as major obstacles to its practical exploitation, as they impact the coherence and performance of the utilized quantum devices. Significant efforts have been dedicated to mitigate and control their negative effects on these devices. This Perspective takes a different approach, aiming to harness the potential of noise and dissipation instead of combatting them. Surprisingly, it is shown that these seemingly detrimental factors can provide substantial advantages in the operation of quantum machine learning algorithms under certain circumstances. Exploring and understanding the implications of adapting quantum machine learning algorithms to open quantum systems opens up pathways for devising strategies that effectively leverage noise and dissipation. The recent works analyzed in this Perspective represent only initial steps towards uncovering other potential hidden benefits that dissipation and noise may offer. As exploration in this field continues, significant discoveries are anticipated that could reshape the future of quantum computing.
\end{abstract}


	\section{Introduction}
	
	Quantum machine learning~\cite{BiamonteNature,LamataReview,Schuld2021,MelnikovReview,LamataReview2} is a rapidly advancing field within quantum technologies, aiming to leverage quantum devices to perform machine learning computations more efficiently. While a definitive demonstration of this capability is yet to be achieved, notable progress is being made in both theoretical and experimental aspects. 
	
	\hfill
	
	One critical aspect affecting quantum systems involved in machine learning protocols is their interaction with the external environment, leading to decoherence. Decoherence is the primary cause of faulty behavior in controllable quantum systems. As quantum machine learning protocols are intended to be implemented in quantum devices, which are inherently susceptible to decoherence, a comprehensive study of quantum machine learning algorithms should account for the impact of the quantum environment, resulting in coherence loss within the system. However, many existing quantum machine learning algorithms in the literature are purely mathematical in nature and do not explore the performance implications of real quantum implementations, which will inevitably be affected by noise and decoherence. Hence, it is crucial to address the influence of the quantum environment and incorporate noise considerations when developing and evaluating quantum machine learning algorithms.
	
	\hfill
	
	The field of open quantum systems~\cite{BreuerPetruccione2003,Rivas_Huelga} investigates the coherence properties of a quantum system when it interacts with an external quantum system or environment, leading to entanglement between them. This field has reached a mature stage, with well-established mathematical formulations involving concepts like Lindblad master equations and density matrices.  Consequently, it was natural for researchers to investigate the application of this formalism in the context of quantum machine learning to understand the impact of decoherence and dissipation on these quantum algorithms.  In this Perspective, rather than focusing on the typical analysis of the negative effects that the interaction with the environment can have on quantum machine learning and exploring ways to mitigate or control them, our main emphasis is on studying the potential benefits that can arise from such interaction. 
	
	\hfill
	
	The structure of the remainder of this work is as follows. In Section~\ref{QML}, we provide a brief review of the topic of quantum machine learning. In Section~\ref{OQS}, we introduce some basic concepts of open quantum systems that will be relevant for the rest of the paper. In Section~\ref{QMLOQS}, we analyze three examples where the interaction with the environment can lead to improvements in the performance of specific quantum machine learning algorithms. Finally, in Section~\ref{Conclusions}, we present a summary and conclusions of the aspects covered in this work.
	
	\section{Quantum Machine Learning}
	\label{QML}
	In the past few years, the field of quantum machine learning~\cite{BiamonteNature,LamataReview,Schuld2021,MelnikovReview,LamataReview2} has significantly grown, producing several important results combining quantum devices with machine learning calculations. The main aim in this field is to accelerate machine learning calculations via employing the speedups produced by genuine quantum properties such as entanglement and superposition. Some important theoretical results and implementations have been produced, including, e.g., linear solvers of equations~\cite{PhysRevLett.103.150502}, quantum principal component analyses~\cite{NatPhysQPCA}, quantum support vector machines~\cite{PRLettQSVM}, quantum annealers~\cite{ReviewQAnnealing}, variational quantum eigensolvers~\cite{ReviewQEigensolver}, quantum Boltzmann machines~\cite{amin2018quantum, kieferova2017tomography}, quantum reinforcement learning~\cite{Dong2008,Paparo2014,Dunjko2016,Bukov2018,Bukov_Day2018,Fosel2018,Liu2022,Albarran_Arriagada_2020}, quantum memristors~\cite{pfeiffer2016quantum}, quantum feature spaces and kernels~\cite{QKernels}, and quantum generative adversarial networks~\cite{QGAN}.  Some of these have speedups relying on the quantum phase estimation algorithm, others are based on Grover search, and others obtain heuristic gains when resources are limited. Even if it is hard to rigorously prove a quantum speedup with respect to any classical machine learning protocol, there is hope inside the quantum machine learning community that this may be one of the areas inside quantum technologies which may have useful applications in industry and society at a nearer time.

\hfill
	
	Some of the advantages of using quantum systems for machine learning tasks arise from the fact that quantum mechanics is well described by linear algebra, which is a common framework in machine learning protocols, at least for some of the subroutines, such as computing distances. In other cases, the speedups may come from combining classical and quantum protocols in a hybrid approach, which can mitigate the exponential explosion characteristic of machine learning calculations through quantum parallelism. Experimental implementations have shown speedups in comparison to classical or other quantum algorithms, particularly on noisy intermediate-scale quantum (NISQ) devices~\cite{QGAN,QKernels,QReinfHefei,QReinfVienna,QMemristVienna,QAdvGoogle}. These studies often explore the advantages of quantum technologies over classical machine learning.  However, the effect of dissipation caused by the surrounding environment on quantum machine learning has not yet been thoroughly analyzed. Real quantum experiments invariably involve working at non-zero temperatures, leading to some level of dissipation. Initial insights into this issue can be found in~\cite{Nguyen2020,Lu2022,OliveraAtencio2023,Domingo2023}.
	
	\hfill
	
	One of the pioneering examples of a quantum machine learning algorithm studied in the presence of dissipation is the quantum reinforcement learning protocol introduced in~\cite{PhysRevA.98.042315,Albarran_Arriagada_2020}. In this protocol, a quantum agent seeks to learn an unknown quantum state provided by the artificial intelligence (AI) environment (not to be confused with the quantum environment producing decoherence), which supplies several copies of it~\cite{PhysRevA.98.042315}. The agent couples with the unknown state and performs measurements on each copy. Depending on the measurement outcome, it either takes no action (exploitation) or applies a random unitary operation (exploration). As the agent approaches the unknown state, the random unitary operations gradually approximate the actual unitary operation. Consequently, the exploration regime diminishes, while the exploitation regime becomes more prominent. The system eventually converges to the unknown state with high fidelity, as experimentally demonstrated in~\cite{QReinfHefei}.
	Subsequently, the protocol was extended to handle unknown eigenstates of an operator, rather than unknown states~\cite{Albarran_Arriagada_2020}. In~\cite{OliveraAtencio2023}, the same protocol was analyzed under the influence of a dissipative bath, using a master equation formalism. The findings suggest that dissipation does not significantly alter the protocol's performance below a certain threshold; in some cases, it can even be beneficial. More details on this work, along with other papers in the literature exploring quantum machine learning protocols and the potential benefits of their interaction with the environment, will be discussed in Section~\ref{QMLOQS}.

\section{Some Basic Notions About Open Quantum Systems}
\label{OQS}
	
Traditionally, the time evolution of quantum systems has been described through a unitary transformation that connects the states of the system at two points in time. If the state of the system at time $t_0$ is represented by the density operator $\rho(t_0)$, then the state at time $t$ is described by the density operator 
\begin{equation}
\rho(t) = U(t,t_0)\rho(t_0)U^\dagger(t,t_0),
\label{Unitary}
\end{equation} 
where $U(t,t_0)$ is a unitary operator known as the time evolution operator from time $t_0$ to time $t$ (see, e.g.,~\cite{Sakurai_Napolitano}). However, this description becomes excessively restrictive when attempting to analyze the time evolution of an open quantum system, i.e., a subsystem within a composite quantum system. Nevertheless, this scenario is the most common in practical terms, as it is nearly impossible experimentally to maintain complete isolation of the quantum system of interest from its surrounding environment. 

\hfill
	
A common approach to analyze the time evolution of an open quantum system (see, e.g.,~\cite{BreuerPetruccione2003}), which will be followed here,  is to consider that the system of interest, S, along with its environment, E, constitutes a closed composite quantum system, C, which evolves according to the unitary time evolution operator $U_{\mathrm{C}}(t,t_0)$.
Therefore, if the density operator of the composite system C at an initial time $t_0$ is $\rho_{\mathrm{C}}(t_0)$, the density operator at time $t$ is given by $\rho_{\mathrm{C}}(t)=U_{\mathrm{C}}(t,t_0)\rho_{\mathrm{C}}(t_0)U_{\mathrm{C}}^{\dagger}(t,t_0)$. The quantum states of the subsystems S and E at time $t$ are represented by their respective reduced density operators denoted by $\rho_{\mathrm{S}}(t)$ and $\rho_{\mathrm{E}}(t)$. These operators contain all the relevant statistical information for potential measurements performed on each subsystem.  
To obtain $\rho_{\mathrm{S}}(t)$ and $\rho_{\mathrm{E}}(t)$, one simply needs to calculate the partial trace of $\rho_{\mathrm{C}}(t)$ with respect to the degrees of freedom of E and S, respectively, i.e., $\rho_{\mathrm{S}}(t)=\mathrm{Tr}_{\mathrm{E}}\left[\rho_{\mathrm{C}}(t)\right]$ and $\rho_{\mathrm{E}}(t)=\mathrm{Tr}_{\mathrm{S}}\left[\rho_{\mathrm{C}}(t)\right]$.  Furthermore, assuming, as is customary, that at the initial time the density operator of C is separable, i.e., that $\rho_{\mathrm{C}}(t_0)=\rho_{\mathrm{S}}(t_0) \otimes \rho_{\mathrm{E}}(t_0)$,   it follows that
\begin{equation}
\rho_{\mathrm{S}}(t)=\mathrm{Tr}_{\mathrm{E}}[U_{\mathrm{C}}(t,t_0)\rho_{\mathrm{S}}(t_0) \otimes \rho_{\mathrm{E}}(t_0)U_{\mathrm{C}}^{\dagger}(t,t_0)].
\end{equation}

\hfill

If the reduced density operator $\rho_E(t_0)$ is taken to be equal to a certain fixed value $\rho_{E,0}$ regardless of the value of $t_0$, the above expression establishes a mapping  $\mathcal{T}_{t,t_0}\left[\rho_{\mathrm{S}}(t_0)\right]\coloneqq\mathrm{Tr}_{\mathrm{E}}[U_{\mathrm{C}}(t,t_0) \rho_{\mathrm{S}}(t_0) \otimes \rho_{\mathrm{E},0}U_{\mathrm{C}}^{\dagger}(t,t_0)]$ 
that assigns to each reduced density operator of S at $t_0$ the corresponding density operator at time $t$. From the definition of partial trace, it can be easily proven that the mapping  $\mathcal{T}_{t,t_0}$ can be expressed in the form~\cite{NielsenChuang2000,BreuerPetruccione2003} 
\begin{equation}
\mathcal{T}_{t,t_0}\left[\rho_{\mathrm{S}}(t_0)\right]=\sum_{\alpha,\beta}K_{\alpha,\beta}(t,t_0)\rho_{\mathrm{S}}(t_0) K_{\alpha,\beta}^{\dagger}(t,t_0), 
\label{krausdec}
\end{equation}
where $K_{\alpha,\beta}(t,t_0)$ are operators acting on the Hilbert space of S, known as Kraus operators~\cite{Kraus_1983}. The explicit form of these operators is 
\begin{equation}
\label{Kraus}
K_{\alpha,\beta}(t,t_0)=\lambda_{\beta}^{1/2} \sum_{l,m}\mel{u_l,\psi_{\alpha}}{U_{\mathrm{C}}(t,t_0) }{u_m,\psi_{\beta}}\ketbra{u_l}{u_m},
\end{equation}
where $\{\ket{u_l}\}$  is an arbitrary orthonormal basis of the Hilbert space of S, $\lambda_{\alpha}$ and $\ket{\psi_{\alpha}}$ are, respectively, the eigenvalues and corresponding eigenvectors of  $\rho_{\mathrm{E},0}$, and we have introduced the shorthand notation $\ket{u_l,\psi_{\alpha}}\coloneqq\ket{u_l}\otimes\ket{\psi_{\alpha}}$. From Equation~(\ref{Kraus}), it is easy to verify that the Kraus operators satisfy the condition 
\begin{equation}
\sum_{\alpha,\beta}K_{\alpha,\beta}^{\dagger}(t,t_0)K_{\alpha,\beta}(t,t_0)=I_{\mathrm{S}},
\label{relation}
\end{equation}
where $I_{\mathrm{S}}$ is the identity operator of the Hilbert space of S. Note that unitary evolution is a particular case of the previous scenario in which only one Kraus operator is present. This case would occur, for example, if there were no interaction between the system S and the environment E, such that $U_{\mathrm{C}}(t,t_0)=U_{\mathrm{S}}(t,t_0)\otimes U_{\mathrm{E}}(t,t_0)$, where $U_{\mathrm{S}}(t,t_0)$ and $U_{\mathrm{E}}(t,t_0)$ are the time evolution operators from time $t_0$ to $t$ for the isolated systems S and E, respectively.

\hfill

The equation governing the temporal evolution of the density operator of a closed quantum system is the Liouville-von Neumann equation
\begin{equation}
\dot{\rho}(t)=-\frac{i}{\hbar}[H(t),\rho(t)],
\label{LVN}
\end{equation}
where an overdot symbolizes a derivative with respect to time, $H(t)$ represents the Hamiltonian of the system, and the square brackets denote the commutator of the operators enclosed within. This equation, which is equivalent to the Schrödinger equation when the system's state is pure, is a direct consequence of the fact that the time evolution operator satisfies the property $U(t,t_0)=U(t,t_1)U(t_1,t_0)$ for all $t, t_1, t_0 \in \mathbb{R}$.  In fact, if we formally define the system's Hamiltonian through the limit $H(t)=i \hbar \lim_{t_1\to t}\partial U(t,t_1)/\partial t$, it is straightforward to see that the Liouville-von Neumann equation~(\ref{LVN}) follows directly from the aforementioned property and  Equation~(\ref{Unitary}).

\hfill

In the case of open quantum systems, it is easy to verify that the property $\mathcal{T}_{t,t_0}=\mathcal{T}_{t,t_1}\mathcal{T}_{t_1,t_0}$ is generally not valid. Indeed, considering that $U_{\mathrm{C}}(t,t_0)=U_{\mathrm{C}}(t,t_1)U_{\mathrm{C}}(t_1,t_0)$, it can be shown that
\begin{equation}
\mathcal{T}_{t,t_0}\left[\rho_{\mathrm{S}}(t_0)\right]=\mathrm{Tr}_{\mathrm{E}}[U_{\mathrm{C}}(t,t_1)\rho_{\mathrm{S}}(t_1)\otimes \rho_{\mathrm{E}}(t_1)U_{\mathrm{C}}^{\dagger}(t,t_1)]+\mathrm{Tr}_{\mathrm{E}}[U_{\mathrm{C}}(t,t_1) \delta\rho_{\mathrm{C}}(t_1) U_{\mathrm{C}}^{\dagger}(t,t_1)],
\label{corr}
\end{equation}
where $\delta\rho_{\mathrm{C}}(t_1)=\rho_{\mathrm{C}}(t_1)- \rho_{\mathrm{S}}(t_1)\otimes \rho_{\mathrm{E}}(t_1)$ quantifies how different $\rho_{\mathrm{C}}(t_1)$ is from a separate state. Since $\rho_{\mathrm{S}}(t_1)=\mathcal{T}_{t_1,t_0}\left[\rho_{\mathrm{S}}(t_0)\right]$, Equation~(\ref{corr}) shows that in general $\mathcal{T}_{t,t_0}\left[\rho_{\mathrm{S}}(t_0)\right]\neq\mathcal{T}_{t,t_1}\mathcal{T}_{t_1,t_0}\left[\rho_{\mathrm{S}}(t_0)\right]$. For $\mathcal{T}_{t,t_0}\left[\rho_{\mathrm{S}}(t_0)\right]$ and $\mathcal{T}_{t,t_1}\mathcal{T}_{t_1,t_0}\left[\rho_{\mathrm{S}}(t_0)\right]$ to be at least approximately equal, two conditions must be met. Firstly, the second term on the right-hand side of Equation~(\ref{corr}) must be negligible compared to the first term. This implies that the density operator of C should remain approximately separable over time, which in turn implies that the interaction between S and E must be sufficiently weak. Secondly, the operator $\rho_{\mathrm{E}}(t_1)$ appearing in the first term on the right-hand side of Equation~(\ref{corr}) must be approximately equal to $\rho_{\mathrm{E},0}$. This, in turn, implies that E must hardly be affected by S and must be in a time-independent steady state. This can be achieved, for example, if E acts as a thermal bath for S in a thermodynamic equilibrium state at a certain temperature $T$. 

\hfill

Using Equation~(\ref{krausdec}) and assuming that the condition $\mathcal{T}_{t,t_0}=\mathcal{T}_{t,t_1}\mathcal{T}_{t_1,t_0}$ is satisfied, at least approximately, it can be shown (see, e.g.,~\cite{BreuerPetruccione2003}) that the reduced density operator of S fulfills the evolution equation
\begin{equation}
	\dot{\rho}_{\mathrm{S}}(t)=-\frac{i}{\hbar}\left[H_{\mathrm{S}}(t),\rho_{\mathrm{S}}(t)\right]+\sum_{j=1}^{N^2-1}\gamma_j(t)\left[L_j(t)\rho_{\mathrm{S}}(t)L_j^{\dagger}(t)-\frac{1}{2}\left\{L_j^{\dagger}(t)L_j(t),\rho_{\mathrm{S}}(t)\right\}\right],
	\label{Lindblad_eq}
\end{equation}
where $H_{\mathrm{S}}(t)$ is an operator playing the role of the Hamiltonian for S but which can also contain information about the environment E, $N$ is the dimension of the Hilbert space of S, $\{\gamma_j(t)\}_{j=1,\dots,N^2-1}$ is a set of nonnegative functions with dimensions of frequency, $\{L_j(t)\}_{j=1,\dots,N^2-1}$ is a set of dimensionless operators known as Lindblad operators or quantum jump operators, and the braces denote the anticommutator of the operators enclosed within.  Equation~(\ref{Lindblad_eq})  is known as the Gorini-Kossakowski-Sudarshan-Lindblad (GKSL) equation or simply as the Lindblad master equation~\cite{LindbladCMP76,GoriniJMP76}. The first term on the right-hand side of Equation~(\ref{Lindblad_eq}) accounts for the unitary or coherent evolution of the system, whereas the second term describes the dissipative aspect of the dynamics. In the typical case where the time evolution operator $U_{\mathrm{C}}(t,t_0)$ is invariant under temporal translations, i.e., $U_{\mathrm{C}}(t+\tau,t_0+\tau)=U_{\mathrm{C}}(t,t_0)$ $\forall t, t_0, \tau\in \mathbb{R}$, the Kraus operators $K_{\alpha,\beta}(t,t_0)$ depend solely on the time difference $t-t_0$, and $H(t)$, $L_j(t)$, and $\gamma_j(t)$ become constants, resulting in a time-independent Lindblad master equation. To obtain explicit expressions for $H(t)$, $L_j(t)$, and $\gamma_j(t)$, it is necessary to start from a specific microscopic model in which the Hamiltonian of the composite system C is provided. The usual methods used to derive Lindblad-type equations from microscopic models, as well as the required approximations, can be found in~\cite{BreuerPetruccione2003, Rivas_Huelga}.

\section{Dissipative Quantum Machine Learning}	
\label{QMLOQS}

Decoherence, arising from the interaction between a quantum system and its environment, is often considered the most formidable obstacle in quantum information science.  The resulting dissipation tends to erode the intriguing quantum phenomena that underlie the potential of quantum computation. To mitigate these adverse effects, various approaches have been developed, including schemes that are robust against them or control mechanisms to counteract their impact (see, e.g.,~\cite{Nguyen2020,Zeng2020}). Although the common perception is that dissipation's effect is negative, it has been demonstrated that in certain cases, dissipation can be leveraged as a valuable resource for universal quantum computation, giving rise to the concept of dissipative quantum computing~\cite{Verstraete2009,Diehl2008,Kraus2008}. The beneficial effects of noise and dissipation have also been observed in other classical and quantum phenomena, such as stochastic resonance~\cite{RevModPhys.70.223,PhysRevLett.91.210601,10.1063/1.1858671} and stochastic synchronization~\cite{10.1063/1.1500497,PhysRevE.71.011101,PhysRevLett.97.210601,PhysRevE.104.064204}.

\hfill

Drawing an analogy, one might ponder the possibility of harnessing dissipation and noise to improve the performance of specific quantum machine learning algorithms. In this section, we will explore this potential and present a series of illustrative examples. Our focus will be specifically on noise and dissipation effects from the surrounding environment, while excluding other works that primarily address stochasticity introduced to the algorithm through factors such as a finite number of measurements or zero-average fluctuations encountered in real experiments (see, e.g.,~\cite{Liu2023stochastic}).  It is important to note that the investigation of noise and dissipation benefits for quantum machine learning is still in its early stages, and the number of published papers exploring this area remains limited.

\subsection{Applications of Dissipative Two-Qubit Systems in Quantum Machine Learning}

\begin{figure}
	\centering
	\includegraphics[width=.5\linewidth]{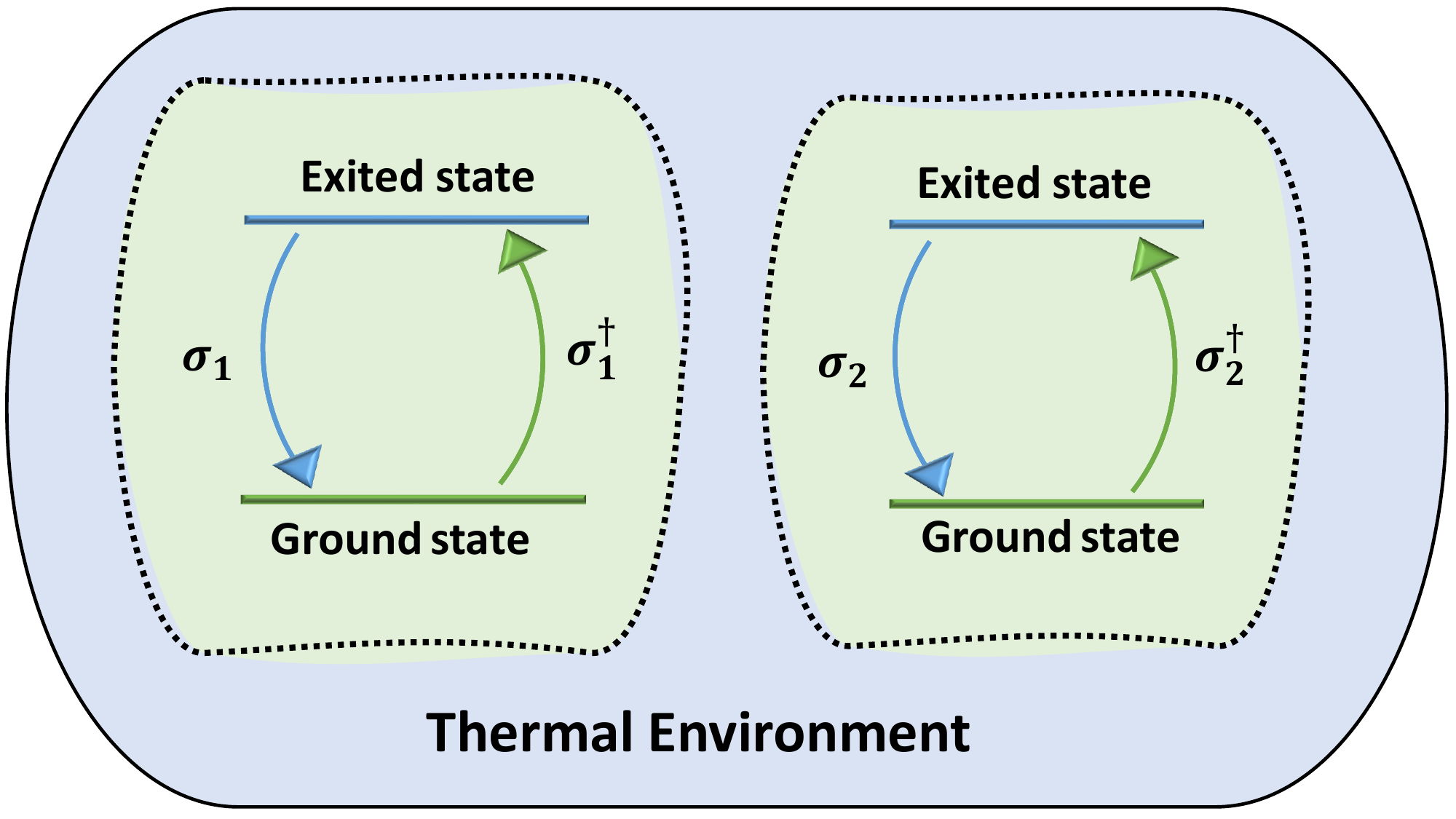}
	\caption{Sketch depicting a two-qubit system undergoing dissipative dynamics through interaction with a common thermal environment.}
	\label{DTQS}
\end{figure}

Several studies~\cite{Ghasemian2021,Ghasemian2022,Ghasemian2023} have explored the possibility of employing a system consisting of two non-interacting qubits coupled to a common thermal bath (see \textbf{Figure~\ref{DTQS}}) as a fundamental unit for quantum machine learning tasks. The evolution of the reduced density operator for the two-qubit system, denoted as $\rho(t)$ (hereafter, we will drop the subscript S to refer to this operator), can be described by the Lindblad master equation~\cite{Ghasemian2020,Ghasemian2021,Ghasemian2022,Ghasemian2023}
\begin{equation}
	\dot{\rho}(t)=\Gamma\sum_{j=1}^2\left[ L_j \rho(t) L_j^{\dagger}-\frac{1}{2}\{L_j^{\dagger}L_j,\rho(t)\}\right],
	\label{LindbladTQB}
\end{equation}
with the Lindblad operators $L_1=\sqrt{\bar{n}+1}(\sigma_1+\sigma_2)$ and $L_2=\sqrt{\bar{n}}(\sigma_1+\sigma_2)^{\dagger}$, where $\sigma_1$ and $\sigma_2$ are the lowering operators of qubit $1$ and $2$, respectively, $\bar{n}$ represents the mean number of thermal excitations corresponding to the global environment, and $\Gamma$ is the spontaneous emission rate. Further information about the derivation of these types of equations can be found in~\cite{Santos2014}. In~\cite{Ghasemian2023}, the case in which the two-qubit system interacts with a squeezed vacuum field reservoir~\cite{Hernandez2008} has also been considered. In this case, the Lindblad master equation that describes the evolution of the density operator of the two qubits is
\begin{equation}
	\dot{\rho}(t)=\Gamma \left[L \rho(t) L^{\dagger}-\frac{1}{2}\{L^{\dagger}L,\rho(t)\}\right],
	\label{Lindbladsqueezed}
\end{equation}
with the Lindblad operator $L=\cosh(r)(\sigma_1+\sigma_2)-\sinh(r)e^{i \psi}(\sigma_1+\sigma_2)^{\dagger}$, where $r$ is the squeezing parameter and $\psi$ the squeezing angle~\cite{Hernandez2008}.

\hfill

From a mathematical perspective, the Lindblad master equations~(\ref{LindbladTQB}) and (\ref{Lindbladsqueezed}) are nothing but a system of first-order linear differential equations with constant coefficients. Given a certain initial condition, $\rho(0)$, this system of differential equations can be solved using standard methods to obtain the value of the density operator of the two-qubit system at any time $t$. By taking the long-time limit of this solution, one can establish a mapping that assigns to each initial condition $\rho(0)$ its corresponding steady state $\rho(\infty)$. The quantum operation or noisy quantum channel corresponding to this mapping can be expressed in the form~\cite{NielsenChuang2000} 
\begin{equation}
	\rho(\infty)=\sum_{j=1}^4 K_j \rho(0) K_j^{\dagger},
	\label{Kraus2}
\end{equation}
where $K_j$ are certain Kraus operators that can be determined by explicitly solving Equation~(\ref{LindbladTQB}) or (\ref{Lindbladsqueezed}). Explicit expressions for these Kraus operators in the case of a thermal bath at zero temperature (i.e., for $\bar{n}=0$) can be found in~\cite{Ghasemian2021,Ghasemian2022}.

\hfill

A detailed analysis of the results obtained from the noisy quantum channel~(\ref{Kraus2}) for the case of a thermal bath [Equation~(\ref{LindbladTQB})] reveals~\cite{Ghasemian2021} that the system retains residual coherence even in the long-time limit, and that it is possible to generate steady entangled states, including Werner-like states~\cite{Werner1989} and maximally entangled mixed states~\cite{Ishizaka2000}. Additionally, since the obtained steady states depend on the chosen initial state, one can manipulate the initial state to construct robust Bell-like states~\cite{Ghasemian2021}.  In the case of the squeezed vacuum field reservoir, the solution of Equation~(\ref{Lindbladsqueezed}) in the small squeezing regime gives rise to the so-called two-qubit X-states~\cite{Ghasemian2023,Yu2007}, which generalize many families of entangled two-qubit states, such as Bell states~\cite{NielsenChuang2000}, Werner states, isotropic states~\cite{Horodecki1999}, and maximally entangled mixed states. 

\hfill

In the literature, several practical applications of the previously described dissipative two-qubit systems to the field of quantum machine learning have been proposed. One such application involves using the noisy quantum channel~(\ref{Kraus2}) to address binary classification problems~\cite{dudaHart1973}. The matrix elements of the initial density operator $\rho(0)$ in a specific basis encode the input attributes, while those of the steady-state $\rho(\infty)$ encode the target attributes. To optimize the classification task, the parameters of the reservoir, such as the temperature or the mean number of thermal excitations $\bar{n}$, for a thermal bath, and the squeezing parameters $r$ and $\psi$, for a squeezed vacuum field reservoir, can be tuned. To establish the classification model, a training set with records containing known class labels and a test set with records containing unknown class labels are required. The training set is used to construct the classification model, which is then applied to the test set for evaluation. This possibility has been explored in~\cite{Ghasemian2021}, where a two-qubit system coupled to a common thermal bath was utilized to classify vertebrates into two distinct groups: mammals and non-mammals. Despite its simplicity, the proposed method demonstrated high accuracy in solving this binary classification problem without requiring any iterative procedures. Moreover, the quantum classifier based on the dissipative two-qubit system outperformed classical algorithms, including the decision tree classifier. As another application, dissipative two-qubit quantum systems have also been proposed as building blocks for the development of quantum neural networks capable of performing classification tasks~\cite{Ghasemian2022,Ghasemian2023}. These results suggest the promising potential of these dissipative systems for efficient and effective classification tasks in the realm of quantum machine learning.

\subsection{Dissipative Quantum Reinforcement Learning}

The possibility of harnessing dissipation to perform quantum machine learning tasks more efficiently has also been explored in the context of quantum reinforcement learning. Specifically, in~\cite{OliveraAtencio2023} the effect of thermal dissipation on a quantum reinforcement learning algorithm has been analyzed, and it has been found that, under certain circumstances, the algorithm can perform better in the presence of dissipation. To understand the implications of these results, we will proceed to briefly describe the studied dissipative quantum algorithm, as well as the benefits that dissipation can bring to its operation.

\hfill

The quantum reinforcement learning algorithm considered in~\cite{OliveraAtencio2023}  is an adaptation to the presence of dissipation of a non-dissipative quantum algorithm analyzed in~\cite{Albarran_Arriagada_2020}. In the dissipative algorithm of~\cite{OliveraAtencio2023}, the agent A is a manipulable qubit that interacts with an AI environment E as well as with a thermal bath B at a finite temperature $T$. The environment E is characterized by an unknown Hamiltonian 
\begin{equation}
H=\frac{\hbar \omega}{2}(\ketbra{+}{+}-\ketbra{-}{-}), 
\end{equation}
where $\omega$ is a positive constant with frequency dimensions, and $\{\ket{+},\ket{-}\}$ are the eigenvectors of $H$. The combined action of E and B on A is described by the Lindblad master equation
\begin{equation}
	\label{Lindblad_TL}
	\dot{\rho}(t)= -\frac{i}{\hbar} [H,\rho(t)]+\sum_{j=\pm} \Gamma_j\left[\tilde{\sigma}_j^{\dagger}\rho(t)\tilde{\sigma}_j-\frac{1}{2}\{\tilde{\sigma}_j\tilde{\sigma}_j^{\dagger},\rho(t)\}\right],
\end{equation}
where $\rho(t)$ is the density operator representing the state of A at time $t$, $\smash{\tilde{\sigma}_-=\ketbra{-}{+}=\tilde{\sigma}_+^{\dagger}}$ is a Lindblad operator that induces dissipative decay from the excited state $\ket{+}$ to the ground state $\ket{-}$,  and $\Gamma_{\pm}=\Gamma_0 e^{\pm \hbar \omega/(2k_B T)}\csch\left[\hbar \omega/(2k_B T)\right] /2$, with $\Gamma_0$ the decay rate from the excited state to the ground state at zero temperature. Given a certain initial condition $\rho(0)$, the solution of Equation~(\ref{Lindblad_TL}) at an arbitrary time $\tau$ can be expressed as
\begin{equation}
	\label{channel1}
	\rho(\tau)=	\mathcal{E}[\rho(0)]\equiv \sum_{j=0}^3 U E_j \rho_0 E_j^{\dagger} U^{\dagger} ,
\end{equation}
where $U=e^{-i \tau H/\hbar}$ and $\{E_0,E_1,E_2,E_3\}$ are the Kraus operators for the generalized amplitude damping channel~\cite{NielsenChuang2000}.

\hfill

The goal of the algorithm is to extract information or learn from the AI environment E to obtain near-optimal knowledge of the eigenstates $\{\ket{+},\ket{-}\}$. The procedure involves performing a large number of iterations, which will be labeled with a natural number $k$. The state of A in the $k$th iteration is denoted by $\ket{\phi_k}$. Furthermore, to facilitate future operations, we also introduce an exploration parameter denoted as $w_k$, which takes on real values between $0$ and $1$. To construct the state $\ket{\phi_{k+1}}$ from the state $\ket{\phi_k}$ and update the value of $w_{k+1}$ from $w_k$, the following steps are followed:
\begin{itemize}
	\item The state of A, represented by the density operator $\ketbra{\phi_k}{\phi_k}$, is allowed to evolve for a time $\tau$ according to the Lindblad master equation~(\ref{Lindblad_TL}). After this evolution, the state of A becomes $\mathcal{E}(\ketbra{\phi_k}{\phi_k})$.
	\item Information is extracted from the state $\mathcal{E}(\ketbra{\phi_k}{\phi_k})$ by performing a measurement of the observable $M_k=\ketbra{\phi_{\perp,k}}{\phi_{\perp,k}}$, where $\ket{\phi_{\perp,k}}$ is the state orthogonal to $\ket{\phi_k}$. In~\cite{OliveraAtencio2023}, a method is proposed to achieve measuring the same observable $M_1$ in all iterations.
	\item If the measurement outcome is $m_k=0$, the state of A after the measurement is $\ket{\phi_k}$. In this case, the state of A remains unchanged for the next iteration, i.e., $\ket{\phi_{k+1}}=\ket{\phi_k}$. Furthermore, the value of the exploration parameter is updated to $w_{k+1}=r w_{k}$, where $0<r<1$ is a real parameter known as the reward rate.
	\item  If, on the contrary,  the measurement outcome is $m_k=1$,  the state of A after the measurement is $\ket{\phi_{\perp,k}}$, and it is necessary to apply the unitary transformation $\sigma_{x,k}=\ketbra{\phi_k}{\phi_{\perp,k}}+\ketbra{\phi_{\perp,k}}{\phi_k}$ to reconstruct the original state $\ket{\phi_k}$. After this transformation, the state of A in the next iteration is constructed by applying a pseudorandom rotation $R_k$ to the state $\ket{\phi_k}$, i.e., $\ket{\phi_{k+1}}=R_k \ket{\phi_k}$.   To implement the pseudorandom rotation $R_k$, three uniformly distributed random angles $\alpha_{x,k}$, $\alpha_{y,k}$, and $\alpha_{z,k}$ are first generated within the interval $\left[-w_k \pi,w_k \pi\right]$. Based on these angles, $R_k$ is given by the expression $R_k=e^{-i \alpha_{y,k} \sigma_{y,k}/2}e^{-i \alpha_{z,k} \sigma_{z,k}/2}e^{-i \alpha_{x,k} \sigma_{x,k}/2}$, with $\sigma_{y,k}=i(\ketbra{\phi_{\perp,k}}{\phi_k}-\ketbra{\psi_k}{\phi_{\perp,k}})$ and $\sigma_{z,k}=\ketbra{\phi_k}{\phi_k}-\ketbra{\phi_{\perp,k}}{\phi_{\perp,k}}$.
	In addition, the value of the exploration parameter is updated to $w_{k+1}=\min(1,p w_k)$, where $p>1$ is a real parameter called the punishment rate. 
\end{itemize}
Using the algorithm described above, the values of $\ket{\phi_k}$ and $w_k$ for any iteration $k$ can be calculated from the initial values $\ket{\phi_1}$ (which is arbitrary) and $w_1=1$. The protocol is considered to converge if $w_k$ approaches zero as $k$ increases, as this indicates that the pseudorandom rotation $R_k$ is converging towards the identity operator. Moreover, the faster it approaches zero, the faster the convergence of the protocol is. Further details about the Markov decision process~\cite{MDP1,MDP2} associated with this protocol can be found in~\cite{OliveraAtencio2023}.

\hfill

The accuracy of the described protocol after performing $k$ iterations can be quantified in two different ways. If the goal is for the agent to learn a specific eigenstate of the Hamiltonian, either the ground state $\ket{-}$ or the excited state $\ket{+}$, the accuracy can be quantified using the corresponding fidelities $f_{-,k}=\abs{\braket{-}{\phi_k}}$ or $f_{+,k}=\abs{\braket{+}{\phi_k}}$. On the other hand, if the objective is for the agent to learn either of the two eigenstates without specifying which one, the fidelity to the closest eigenstate should be chosen, i.e., $f_k=\max(f_{-,k}, f_{+,k})$. Due to the inherently random nature of the measurement outcomes of $M_k$, as well as the pseudorandomness of the angles $\alpha_{x,k}$, $\alpha_{y,k}$, and $\alpha_{z,k}$, the quantities $w_k$, $f_{\mp,k}$, and $f_k$ are random variables whose values vary from one realization to another. For this reason, in~\cite{OliveraAtencio2023}, their mean values obtained from a large number of realizations $N$ are analyzed, which are denoted as $W_k$, $F_{\mp,k}$, and $F_k$, respectively. 

\hfill

\begin{figure}
	\centering
	\includegraphics[width=.6\linewidth]{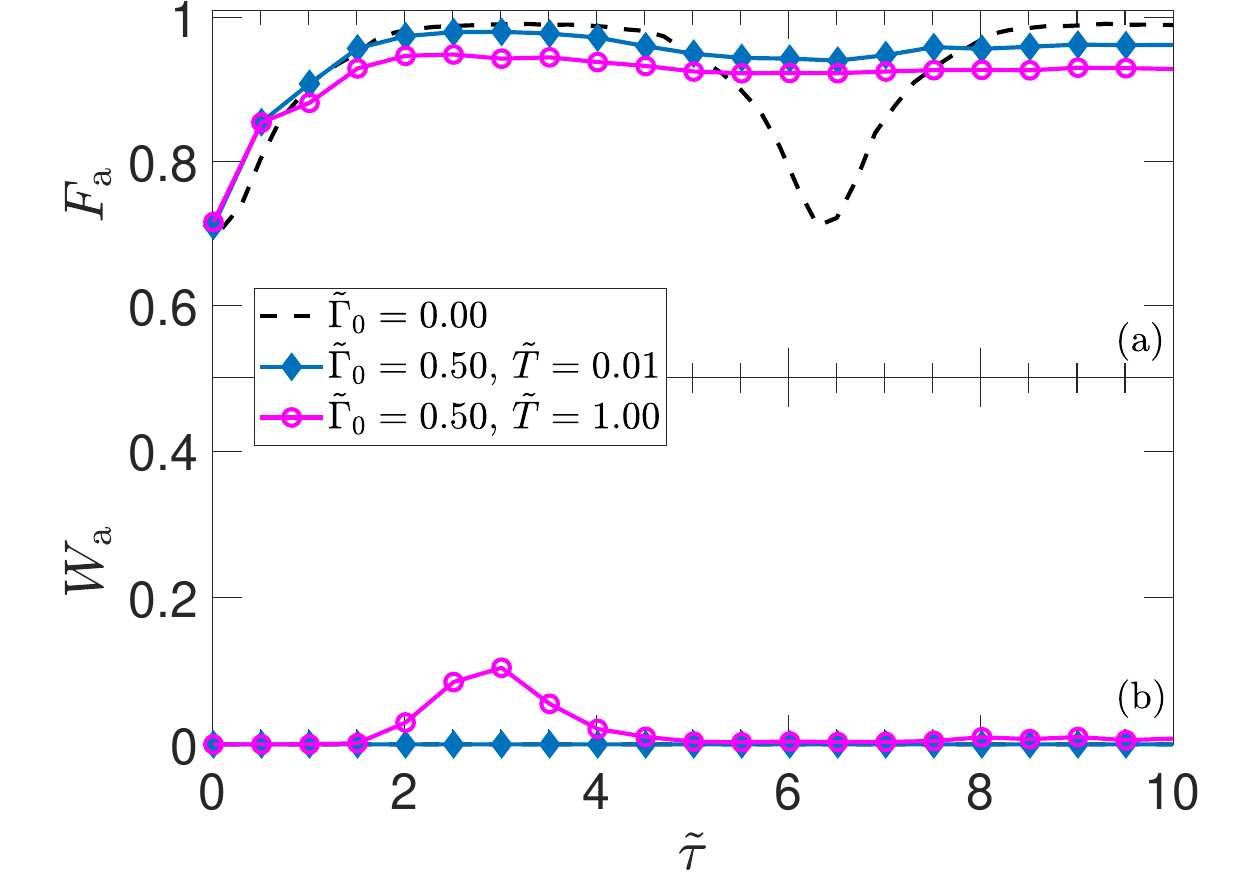}
	\caption{Asymptotic behavior in the large iteration limit (i.e., for $k\gg 1$) of the mean fidelity $F_{\mathrm{a}}$ (a)  and the mean exploration parameter $W_{\mathrm{a}}$ (b) as a function of the dimensionless evolution time $\tilde{\tau}=\omega \tau$  for different values of the dimensionless decay rate at zero temperature $\tilde{\Gamma}_0=\Gamma_0/\omega$ and the dimensionless temperature $\tilde{T}=k_{B}T/(\hbar \omega)$.  The nondissipative case $\tilde\Gamma_0=0$ is represented by black dashed lines, while the dissipative cases are shown by light blue diamond lines ($\tilde{\Gamma}_0=0.5$ and $\tilde{T}=0.01$) and  magenta circled lines ($\tilde{\Gamma}_0=0.5$ and $\tilde{T}=1$). In (b) the black dashed line is indistinguishable from the abscissa axis. The initial state is $\ket{\phi_1}=\ket{0}$ and the remaining parameter values $N = 1000$, $r = 0.9$, and $p = 20/9$.}
	\label{FaWavstau}
\end{figure}

An interesting finding from this analysis is that, at sufficiently low temperatures, dissipation does not always lead to a negative impact on the accuracy of the protocol as quantified by the mean fidelity $F_k$. In fact, there are parameter ranges for which the protocol performs better in the presence of dissipation than in its absence~\cite{OliveraAtencio2023}. To illustrate this fact, in \textbf{Figure~\ref{FaWavstau}}, we have plotted the asymptotic values in the large iteration limit (i.e., for $k\gg 1$) of the mean fidelity $F_k$ and the mean  exploration parameter $W_k$, denoted as $F_{\mathrm{a}}$ and $W_{\mathrm{a}}$, respectively, as a function of the dimensionless evolution time $\tilde{\tau}=\omega \tau$ for different values of the dimensionless decay rate at zero temperature, $\tilde{\Gamma}_0=\Gamma_0/\omega$, and the dimensionless temperature, $\tilde{T}=k_{B}T/(\hbar \omega)$. In this figure, it can be observed that the values of $F_{\mathrm{a}}$ for the dissipative cases are substantially higher than for the nondissipative case for values of $\tilde{\tau}$ around $2\pi$. Moreover, for these values, the protocol converges correctly, as the values of $W_{\mathrm{a}}$ are close to $0$. An explanation for this behavior can be found in~\cite{OliveraAtencio2023}. This finding could be significant for the practical implementation of this protocol, as in some experimental scenarios, the presence of dissipation may be inevitable.

\hfill

\begin{figure}
	\centering
	\includegraphics[width=.6\linewidth]{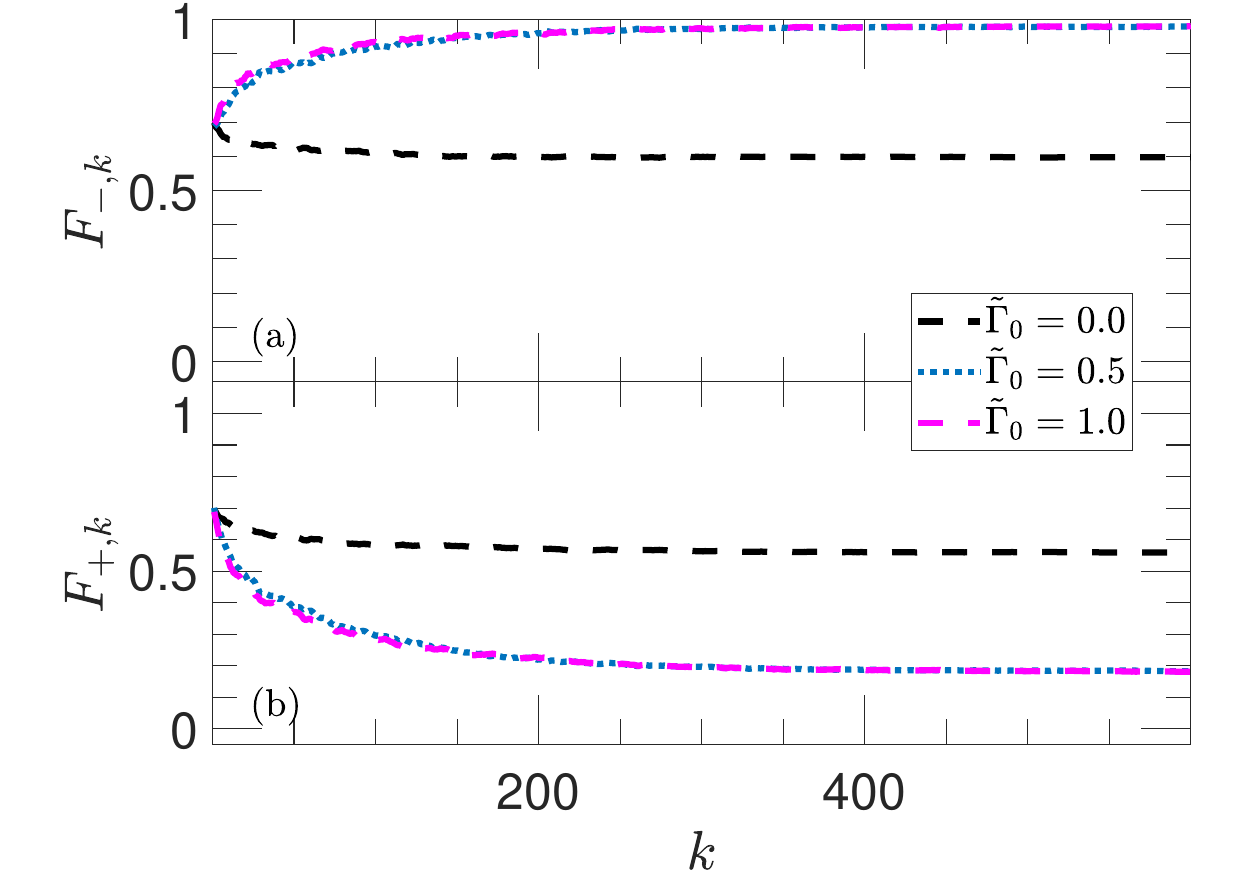}
	\caption{Behavior of the fidelities $F_{-,k}$ (a) and $F_{+,k}$ (b) as a function of the number of iterations $k$. The nondissipative case $\tilde{\Gamma}_0=0$ is represented by black dashed lines, while the dissipative cases are shown by light blue dotted lines ($\tilde{\Gamma}_0=0.5$) and magenta dashed lines ($\tilde{\Gamma}_0=1$). The initial state is $\ket{\phi_1}=\ket{0}$ and the remaining parameter values $\tilde{T}=0.3$, $\tilde{\tau}=1$, $N=1000$, $r=0.9$, and $p=20/9$.}
	\label{FnFpvsk}
\end{figure}

Another quite common scenario in which dissipation plays a particularly positive role is when we are interested in calculating the ground state of the system. In this case, as mentioned earlier, the mean fidelity that should be maximized is $F_{-,k}$ instead of $F_k$. \textbf{Figure~\ref{FnFpvsk}} shows the dependencies of $F_{-,k}$ and $F_{+,k}$ with respect to the number of iterations $k$ in the absence of dissipation, i.e., for $\tilde{\Gamma}_0=0$, represented by black dashed lines. The figure also shows these dependencies when dissipation is present, represented by light blue dotted lines ($\tilde{\Gamma}_0=0.5$) and magenta dashed lines ($\tilde{\Gamma}_0=1$). Figure~\ref{FnFpvsk} clearly shows that, for temperatures not too high, the presence of dissipation significantly affects the values of $F_{-,k}$ and $F_{+,k}$, increasing $F_{-,k}$ and decreasing $F_{+,k}$ compared to the nondissipative case. This is because, for the dissipation and temperature values considered in the figure, the state $\ketbra{-}{-}$ is an approximate fixed point of the channel $\mathcal{E}$ (for more details, see~\cite{OliveraAtencio2023}). As the temperature value increases, this is no longer the case, and the difference becomes less noticeable~\cite{OliveraAtencio2023}).
As a result, if the goal in a real experiment is to calculate the ground state, a certain degree of dissipation can be beneficial, as long as the temperature is not too high.

\hfill

The example we have just analyzed highlights that dissipation can also play a beneficial role in quantum reinforcement learning. However, further investigation is needed to determine whether this assertion holds true for the current protocol when increasing the number of qubits or in other types of quantum reinforcement learning protocols.

\subsection{Harnessing Noise in Quantum Reservoir Computing}

The potential beneficial effects of noise in quantum reservoir computing (QRC)~\cite{Mujal2021} have been recently analyzed in~\cite{Domingo2023}. The concept behind QRC involves employing a Hilbert space as an enhanced feature space for input data. This enhanced feature space, generated through quantum entangling operations, is then utilized to provide input to a classical machine learning model, which performs the desired target prediction. 

\hfill

The problem addressed in~\cite{Domingo2023} involves the prediction of the excited electronic energy $E_1(R)$ of the LiH molecule at different internuclear distances $R$ based on the associated ground state $\ket{\psi_0}_R$ with energy $E_0(R)$. The ground state $\ket{\psi_0}_R$, which can be represented using $n=8$ qubits, is obtained for various values of $R$ through exact diagonalization~\cite{Domingo2022}. The dataset $\{\ket{\psi_0}_R, \Delta E(R)\}$, where $\Delta E(R)=E_1(R)-E_0(R)$ represents the relative excited energy, is divided into training and test sets, with the test set containing $30\%$ of the data to challenge the algorithm for extrapolation to new samples. The training procedure of the algorithm consists of the following steps~\cite{Domingo2023}. Initially, the quantum circuit is prepared with the molecular ground state $\ket{\psi_0}_R$ corresponding to a specific configuration $R$. Then, a noisy quantum circuit with a fixed number of gates is applied to $\ket{\psi_0}_R$. Later, measurements of the local Pauli operators $\{X_0, Z_0, \dots, X_n, Z_n\}$ are performed, where $X_j$ and $Z_j$ denote the Pauli operators $X$ and $Z$ applied to the $j$-th qubit. This process results in the vector $X(R) = (\langle X_0 \rangle, \langle Z_0 \rangle, \dots, \langle X_n \rangle, \langle Z_n \rangle)^T$, which contains the extracted information from the ground state. After obtaining the vector $X(R)$, it is used as input for a classical machine learning algorithm, such as ridge regression, which is a linear model with $L^2$ regularization~\cite{Domingo2022}.

\hfill

Three types of noisy quantum channels~\cite{NielsenChuang2000} are considered in~\cite{Domingo2023}, namely, the amplitude damping channel $\mathcal{E}(\rho)=\sum_{j=0}^1 K_j \rho K_j^{\dagger}$, with Kraus operators 
\begin{equation}
K_0=\ketbra{0}{0}+\sqrt{1-p}\ketbra{1}{1} \quad \mathrm{and} \quad K_1=\sqrt{p}\ketbra{0}{1},
\end{equation}
 the phase damping channel $\mathcal{E}(\rho)=\sum_{j=0}^2 K_j \rho K_j^{\dagger}$, with Kraus operators 
 \begin{equation}
 K_0=\sqrt{1-p}\mathds{1}, \quad K_1=\sqrt{p}\ketbra{0}{0}, \quad \mathrm{and} \quad K_2=\sqrt{p}\ketbra{1}{1},
 \end{equation}
 and the depolarizing channel $\mathcal{E}(\rho)=\sum_{j=0}^3 K_j \rho K_j^{\dagger}$, with Kraus operators 
 \begin{equation}
 K_0=\sqrt{1-p}\mathds{1}, \quad K_1=\sqrt{\frac{p}{3}}X, \quad K_2=\sqrt{\frac{p}{3}} Y, \quad \mathrm{and} \quad K_3=\sqrt{\frac{p}{3}}Z,
 \end{equation}
  where $p$ is the error probability. These models encompass the vast majority of noise types that modern hardware is subjected to. 
  
 \hfill
  
The analysis of the obtained results reveals a surprising finding: under specific conditions, the presence of amplitude damping noise can enhance the performance of QRC. However, in contrast, both the depolarizing and phase damping channels result in poorer outcomes for QRC, with the depolarizing channel showing even worse performance than the phase damping channel. As a consequence, to achieve successful quantum machine learning tasks with quantum reservoirs, addressing depolarizing noise becomes a priority, requiring the implementation of suitable error-correcting methods. In~\cite{Domingo2023}, the authors also propose a possible explanation for these findings based on the distribution of resulting density matrices in the Pauli space after experiencing the different noisy quantum channels. The amplitude damping channel introduces additional non-zero coefficients in the Pauli space, effectively simulating the effect of having more quantum gates in the original circuits, which improves performance. On the contrary, the depolarizing and phase damping channels simply reduce the amplitude of coefficients in the Pauli space, resulting in inferior outcomes. Among these channels, the depolarizing one exhibits the fastest mitigation of these values, which is why its performance is worse than that of the phase damping channel.

\section{Conclusions}
\label{Conclusions}

This Perspective article has shed light on the intriguing intersection between two well-established domains of Quantum Mechanics: quantum machine learning and open quantum systems. By investigating the potential impact of dissipation arising from the interaction between quantum devices and their physical environment, we have uncovered the possibility of beneficial effects on the learning tasks at hand, contrary to the conventional belief of adverse consequences.

\hfill

The recognition of dissipation as a potentially useful resource in quantum machine learning opens up new avenues for the design of strategies that harness noise and dissipation, thereby driving significant advancements in quantum computation. These findings challenge the prevailing notion that dissipation is solely a hindrance and provide a fresh perspective on leveraging quantum properties to enhance learning protocols.

\hfill

The recent works discussed in this Perspective represent initial steps towards understanding the hidden benefits of dissipation in quantum machine learning. As we continue to explore this exciting field, we anticipate the emergence of transformative discoveries that can reshape the future of quantum computing. The integration of quantum machine learning with open quantum systems offers a promising direction for advancing quantum technologies, with potential applications across various fields. Further exploration in this rich and uncharted territory holds the promise of accelerating the development and practical integration of quantum technologies into our daily lives.



\medskip
\textbf{Acknowledgements} \par 
We acknowledge funding by the Junta de Andaluc\'ia and FEDER (P20-00617 and  US-1380840) and by the Spanish Ministry of Science, Innovation, and Universities under grant Nos. PID2019-104002GB-C21 and PID2019-104002GB- C22.
\medskip


\begin{thebibliography}{10}
	\providecommand{\url}[1]{\texttt{#1}}
	\providecommand{\urlprefix}{URL }
	
	\bibitem{BiamonteNature}
	J.~Biamonte, P.~Wittek, N.~Pancotti, P.~Rebentrost, N.~Wiebe, S.~Lloyd,
	\newblock \emph{Nature} \textbf{2017}, \emph{549} 195.
	
	\bibitem{LamataReview}
	L.~Lamata,
	\newblock \emph{Mach. Learn. Sci. Technol.} \textbf{2020}, \emph{1} 033002.
	
	\bibitem{Schuld2021}
	M.~Schuld, F.~Petruccione,
	\newblock \emph{Machine Learning with Quantum Computers},
	\newblock Springer International Publishing, \textbf{2021}.
	
	\bibitem{MelnikovReview}
	A.~Melnikov, M.~Kordzanganeh, A.~Alodjants, R.-K. Lee,
	\newblock \emph{Adv. Phys.: X} \textbf{2023}, \emph{8}, 1 2165452.
	
	\bibitem{LamataReview2}
	L.~Lamata,
	\newblock \emph{Adv. Quantum Technol.} \textbf{2023}, \emph{6}, 7 2300059.
	
	\bibitem{BreuerPetruccione2003}
	H.-P. Breuer, F.~Petruccione,
	\newblock \emph{Theory of Open Quantum Systems},
	\newblock Oxford University Press, Oxford, \textbf{2003}.
	
	\bibitem{Rivas_Huelga}
	A.~Rivas, S.~F. Huelga,
	\newblock \emph{Open Quantum Systems: An Introduction (SpringerBriefs in
		Physics)},
	\newblock Springer, paperback edition, \textbf{2011}.
	
	\bibitem{PhysRevLett.103.150502}
	A.~W. Harrow, A.~Hassidim, S.~Lloyd,
	\newblock \emph{Phys. Rev. Lett.} \textbf{2009}, \emph{103} 150502.
	
	\bibitem{NatPhysQPCA}
	S.~Lloyd, M.~Mohseni, P.~Rebentrost,
	\newblock \emph{Nat. Phys.} \textbf{2014}, \emph{10} 631.
	
	\bibitem{PRLettQSVM}
	P.~Rebentrost, M.~Mohseni, S.~Lloyd,
	\newblock \emph{Phys. Rev. Lett.} \textbf{2014}, \emph{113} 130503.
	
	\bibitem{ReviewQAnnealing}
	R.~K. Nath, H.~Thapliyal, T.~S. Humble,
	\newblock \emph{SN Comp. Sci.} \textbf{2021}, \emph{2} 365.
	
	\bibitem{ReviewQEigensolver}
	J.~R. McClean, J.~Romero, R.~Babbush, A.~Aspuru-Guzik,
	\newblock \emph{New J. Phys.} \textbf{2016}, \emph{18} 023023.
	
	\bibitem{amin2018quantum}
	M.~H. Amin, E.~Andriyash, J.~Rolfe, B.~Kulchytskyy, R.~Melko,
	\newblock \emph{Phys. Rev. X} \textbf{2018}, \emph{8}, 2 021050.
	
	\bibitem{kieferova2017tomography}
	M.~Kieferov{\'a}, N.~Wiebe,
	\newblock \emph{Phys. Rev. A} \textbf{2017}, \emph{96}, 6 062327.
	
	\bibitem{Dong2008}
	D.~Dong, C.~Chen, H.~Li, T.-J. Tarn,
	\newblock \emph{IEEE Transactions on Systems, Man, and Cybernetics, Part B
		(Cybernetics)} \textbf{2008}, \emph{38}, 5 1207.
	
	\bibitem{Paparo2014}
	G.~D. Paparo, V.~Dunjko, A.~Makmal, M.~A. Martin-Delgado, H.~J. Briegel,
	\newblock \emph{Phys. Rev. X} \textbf{2014}, \emph{4} 031002.
	
	\bibitem{Dunjko2016}
	V.~Dunjko, J.~M. Taylor, H.~J. Briegel,
	\newblock \emph{Phys. Rev. Lett.} \textbf{2016}, \emph{117} 130501.
	
	\bibitem{Bukov2018}
	M.~Bukov,
	\newblock \emph{Phys. Rev. B} \textbf{2018}, \emph{98} 224305.
	
	\bibitem{Bukov_Day2018}
	M.~Bukov, A.~G.~R. Day, D.~Sels, P.~Weinberg, A.~Polkovnikov, P.~Mehta,
	\newblock \emph{Phys. Rev. X} \textbf{2018}, \emph{8} 031086.
	
	\bibitem{Fosel2018}
	T.~F\"osel, P.~Tighineanu, T.~Weiss, F.~Marquardt,
	\newblock \emph{Phys. Rev. X} \textbf{2018}, \emph{8} 031084.
	
	\bibitem{Liu2022}
	Y.-P. Liu, Q.-S. Jia, X.~Wang,
	\newblock \emph{IFAC-PapersOnLine} \textbf{2022}, \emph{55}, 11 132, iFAC
	Workshop on Control for Smart Cities CSC 2022.
	
	\bibitem{Albarran_Arriagada_2020}
	F.~Albarr{\'{a}}n-Arriagada, J.~C. Retamal, E.~Solano, L.~Lamata,
	\newblock \emph{Mach. Learn.: Sci. Technol.} \textbf{2020}, \emph{1}, 1 015002.
	
	\bibitem{pfeiffer2016quantum}
	P.~Pfeiffer, I.~Egusquiza, M.~Di~Ventra, M.~Sanz, E.~Solano,
	\newblock \emph{Sci. Rep.} \textbf{2016}, \emph{6}, 1 29507.
	
	\bibitem{QKernels}
	V.~Havlíček, A.~D. Córcoles, K.~Temme, A.~W. Harrow, A.~Kandala, J.~M. Chow,
	J.~M. Gambetta,
	\newblock \emph{Nature} \textbf{2019}, \emph{567} 209.
	
	\bibitem{QGAN}
	L.~Hu, S.-H. Wu, W.~Cai, Y.~Ma, X.~Mu, Y.~Xu, H.~Wang, Y.~Song, D.-L. Deng,
	C.-L. Zou, L.~Sun,
	\newblock \emph{Sci. Adv.} \textbf{2019}, \emph{5}, 1 eaav2761.
	
	\bibitem{QReinfHefei}
	S.~Yu, F.~Albarrán-Arriagada, J.~C. Retamal, Y.-T. Wang, W.~Liu, Z.-J. Ke,
	Y.~Meng, Z.-P. Li, J.-S. Tang, E.~Solano, L.~Lamata, C.-F. Li, G.-C. Guo,
	\newblock \emph{Adv. Quantum Technol.} \textbf{2019}, \emph{2}, 7-8 1800074.
	
	\bibitem{QReinfVienna}
	V.~Saggio, B.~E. Asenbeck, A.~Hamann, T.~Strömberg, P.~Schiansky, V.~Dunjko,
	N.~Friis, N.~C. Harris, M.~Hochberg, D.~Englund, S.~Wölk, H.~J. Briegel,
	P.~Walther,
	\newblock \emph{Nature} \textbf{2021}, \emph{591} 229.
	
	\bibitem{QMemristVienna}
	M.~Spagnolo, J.~Morris, S.~Piacentini, M.~Antesberger, F.~Massa, A.~Crespi,
	F.~Ceccarelli, R.~Osellame, P.~Walther,
	\newblock \emph{Nat. Photon.} \textbf{2022}, \emph{16} 318.
	
	\bibitem{QAdvGoogle}
	H.-Y. Huang, M.~Broughton, J.~Cotler, S.~Chen, J.~Li, M.~Mohseni, H.~Neven,
	R.~Babbush, R.~Kueng, J.~Preskill, J.~R. McClean,
	\newblock \emph{Science} \textbf{2022}, \emph{376}, 6598 1182.
	
	\bibitem{Nguyen2020}
	N.~H. Nguyen, E.~C. Behrman, J.~E. Steck,
	\newblock \emph{Quantum Mach. Intell.} \textbf{2020}, \emph{2} 1.
	
	\bibitem{Lu2022}
	C.~Lu, S.~Kundu, A.~Arunachalam, K.~Basu,
	\newblock In \emph{2022 IEEE 15th Dallas Circuit And System Conference}.
	\textbf{2022} 1--2.
	
	\bibitem{OliveraAtencio2023}
	M.~L. Olivera-Atencio, L.~Lamata, M.~Morillo, J.~Casado-Pascual,
	\newblock \emph{Phys. Rev. E} \textbf{2023}, \emph{108} 014128.
	
	\bibitem{Domingo2023}
	L.~Domingo, G.~Carlo, F.~Borondo,
	\newblock \emph{Sci. Rep.} \textbf{2023}, \emph{13}, 1.
	
	\bibitem{PhysRevA.98.042315}
	F.~Albarr\'an-Arriagada, J.~C. Retamal, E.~Solano, L.~Lamata,
	\newblock \emph{Phys. Rev. A} \textbf{2018}, \emph{98} 042315.
	
	\bibitem{Sakurai_Napolitano}
	J.~J. Sakurai, J.~Napolitano,
	\newblock \emph{Modern Quantum Mechanics},
	\newblock Cambridge University Press, hardcover edition, \textbf{2020}.
	
	\bibitem{NielsenChuang2000}
	M.~A. Nielsen, I.~L. Chuang,
	\newblock \emph{Quantum Computing and Quantum Information},
	\newblock Cambridge University Press, Cambridge, \textbf{2000}.
	
	\bibitem{Kraus_1983}
	K.~Kraus,
	\newblock \emph{States, Effects, and Operations: Fundamental Notions of Quantum
		Theory (Lecture Notes in Physics, 190)},
	\newblock Springer, paperback edition, \textbf{1983}.
	
	\bibitem{LindbladCMP76}
	G.~Lindblad,
	\newblock \emph{Commun. Math. Phys.} \textbf{1976}, \emph{48} 119.
	
	\bibitem{GoriniJMP76}
	V.~Gorini, A.~Kossakowski, E.~C.~G. Sudarshan,
	\newblock \emph{J. Math. Phys.} \textbf{1976}, \emph{17} 821.
	
	\bibitem{Zeng2020}
	Y.~Zeng, J.~Shen, S.~Hou, T.~Gebremariam, C.~Li,
	\newblock \emph{Phys. Lett. A} \textbf{2020}, \emph{384}, 35 126886.
	
	\bibitem{Verstraete2009}
	F.~Verstraete, M.~M. Wolf, J.~I. Cirac,
	\newblock \emph{Nat. Phys.} \textbf{2009}, \emph{5}, 9 633.
	
	\bibitem{Diehl2008}
	S.~Diehl, A.~Micheli, A.~Kantian, B.~Kraus, H.~P. B\"{u}chler, P.~Zoller,
	\newblock \emph{Nat. Phys.} \textbf{2008}, \emph{4}, 11 878.
	
	\bibitem{Kraus2008}
	B.~Kraus, H.~P. B\"{u}chler, S.~Diehl, A.~Kantian, A.~Micheli, P.~Zoller,
	\newblock \emph{Phys. Rev. A} \textbf{2008}, \emph{78}, 4.
	
	\bibitem{RevModPhys.70.223}
	L.~Gammaitoni, P.~H\"anggi, P.~Jung, F.~Marchesoni,
	\newblock \emph{Rev. Mod. Phys.} \textbf{1998}, \emph{70} 223.
	
	\bibitem{PhysRevLett.91.210601}
	J.~Casado-Pascual, J.~G\'omez-Ord\'o\~nez, M.~Morillo, P.~H\"anggi,
	\newblock \emph{Phys. Rev. Lett.} \textbf{2003}, \emph{91} 210601.
	
	\bibitem{10.1063/1.1858671}
	J.~Casado-Pascual, J.~Gómez-Ordóñez, M.~Morillo,
	\newblock \emph{Chaos} \textbf{2005}, \emph{15}, 2 026115.
	
	\bibitem{10.1063/1.1500497}
	J.~A. Freund, L.~Schimansky-Geier, P.~Hänggi,
	\newblock \emph{Chaos} \textbf{2003}, \emph{13}, 1 225.
	
	\bibitem{PhysRevE.71.011101}
	J.~Casado-Pascual, J.~G\'omez-Ord\'o\~nez, M.~Morillo, J.~Lehmann, I.~Goychuk,
	P.~H\"anggi,
	\newblock \emph{Phys. Rev. E} \textbf{2005}, \emph{71} 011101.
	
	\bibitem{PhysRevLett.97.210601}
	I.~Goychuk, J.~Casado-Pascual, M.~Morillo, J.~Lehmann, P.~H\"anggi,
	\newblock \emph{Phys. Rev. Lett.} \textbf{2006}, \emph{97} 210601.
	
	\bibitem{PhysRevE.104.064204}
	M.~L. Olivera-Atencio, M.~Morillo, J.~Casado-Pascual,
	\newblock \emph{Phys. Rev. E} \textbf{2021}, \emph{104} 064204.
	
	\bibitem{Liu2023stochastic}
	J.~Liu, F.~Wilde, A.~A. Mele, L.~Jiang, J.~Eisert,
	\newblock \emph{arXiv:2210.06723v2 [quant-ph]} \textbf{2023}.
	
	\bibitem{Ghasemian2021}
	E.~Ghasemian, M.~K. Tovassoly,
	\newblock \emph{Sci. Rep.} \textbf{2021}, \emph{11} 3554.
	
	\bibitem{Ghasemian2022}
	E.~Ghasemian, M.~K. Tavassoly,
	\newblock \emph{Sci. Rep.} \textbf{2022}, \emph{12}, 1.
	
	\bibitem{Ghasemian2023}
	E.~Ghasemian,
	\newblock \emph{Quantum Mach. Intell.} \textbf{2023}, \emph{5} 13.
	
	\bibitem{Ghasemian2020}
	E.~Ghasemian, M.~K. Tavassoly,
	\newblock \emph{Int. J. Theor. Phys.} \textbf{2020}, \emph{59}, 6 1742.
	
	\bibitem{Santos2014}
	J.~P. Santos, F.~L. Semi{\~{a}}o,
	\newblock \emph{Phys. Rev. A} \textbf{2014}, \emph{89}, 2.
	
	\bibitem{Hernandez2008}
	M.~Hernandez, M.~Orszag,
	\newblock \emph{Phys. Rev. A} \textbf{2008}, \emph{78}, 4.
	
	\bibitem{Werner1989}
	R.~F. Werner,
	\newblock \emph{Phys. Rev. A} \textbf{1989}, \emph{40}, 8 4277.
	
	\bibitem{Ishizaka2000}
	S.~Ishizaka, T.~Hiroshima,
	\newblock \emph{Phys. Rev. A} \textbf{2000}, \emph{62}, 2.
	
	\bibitem{Yu2007}
	T.~Yu, J.~H. Eberly,
	\newblock \emph{Quantum Inf. Comput.} \textbf{2007}, \emph{7}, 5 459–468.
	
	\bibitem{Horodecki1999}
	M.~Horodecki, P.~Horodecki,
	\newblock \emph{Phys. Rev. A} \textbf{1999}, \emph{59}, 6 4206.
	
	\bibitem{dudaHart1973}
	R.~O. Duda, P.~E. Hart,
	\newblock \emph{Pattern Classification and Scene Analysis},
	\newblock John Willey \& Sons, New Yotk, \textbf{1973}.
	
	\bibitem{MDP1}
	M.~L. Puterman,
	\newblock In \emph{Stochastic Models}, volume~2 of \emph{Handbooks in
		Operations Research and Management Science}, 331--434. Elsevier,
	\textbf{1990}.
	
	\bibitem{MDP2}
	M.~L. Puterman,
	\newblock \emph{Markov decision processes: discrete stochastic dynamic
		programming},
	\newblock John Wiley \& Sons, \textbf{2014}.
	
	\bibitem{Mujal2021}
	P.~Mujal, R.~Mart{\'{\i}}nez-Pe{\~{n}}a, J.~Nokkala, J.~Garc{\'{\i}}a-Beni,
	G.~L. Giorgi, M.~C. Soriano, R.~Zambrini,
	\newblock \emph{Adv. Quantum Technol.} \textbf{2021}, \emph{4}, 8 2100027.
	
	\bibitem{Domingo2022}
	L.~Domingo, G.~Carlo, F.~Borondo,
	\newblock \emph{Phys. Rev. E} \textbf{2022}, \emph{106}, 4.
	
\end{thebibliography}
\end{document}